\documentclass[aps, prl, showpacs, twocolumn, superscriptaddress]{revtex4-1}
\usepackage{bm, amsmath, amsfonts, amssymb}
\usepackage{braket}
\usepackage[dvipdfmx]{graphicx}
\usepackage{float}
\usepackage{color}

\usepackage{comment}
\usepackage{xspace}
\usepackage{stmaryrd}
\usepackage{ulem}

\definecolor{dgreen}{rgb}{0.0, 0.5, 0.0}

\newcommand{\ii}{\text{i}}

\begin{document}


\title{Information Retrieval and Criticality in Parity-Time-Symmetric Systems}

\author{Kohei Kawabata}
\affiliation{Department of Physics, University of Tokyo, 7-3-1 Hongo, Bunkyo-ku, Tokyo
113-0033, Japan}

\author{Yuto Ashida}
\affiliation{Department of Physics, University of Tokyo, 7-3-1 Hongo, Bunkyo-ku, Tokyo
113-0033, Japan}

\author{Masahito Ueda}
\affiliation{Department of Physics, University of Tokyo, 7-3-1 Hongo, Bunkyo-ku, Tokyo
113-0033, Japan}
\affiliation{RIKEN Center for Emergent Matter Science (CEMS), Wako, Saitama 351-0198, Japan}

\date{\today}

\begin{abstract}
By investigating information flow between a general parity-time (PT) -symmetric non-Hermitian system and an environment, we find that the complete information retrieval from the environment can be achieved in the PT-unbroken phase, whereas no information can be retrieved in the PT-broken phase. The PT-transition point thus marks the reversible-irreversible criticality of information flow, around which many physical quantities such as the recurrence time and the distinguishability between quantum states exhibit power-law behavior. Moreover, by embedding a PT-symmetric system into a larger Hilbert space so that the entire system obeys unitary dynamics, we reveal that behind the information retrieval lies a hidden entangled partner protected by PT symmetry. Possible experimental situations are also discussed.
\end{abstract}

\maketitle

Non-Hermitian systems with parity-time (PT) symmetry have attracted growing interest over the past two decades \cite{Bender-98, Bender-review}. PT-symmetric systems have two phases: the unbroken phase where the entire eigenspectrum is real, and the broken phase where some eigenvalues form complex conjugate pairs. Between the two phases lies an exceptional point where an unconventional phase transition occurs \cite{Heiss-12}. Several unique properties of PT-symmetric systems have been predicted and observed in classical systems where gain and loss are balanced \cite{El-Ganainy-07, Makris-08, Guo-09, Ruter-10, Longhi-09, Wimmer-15, Bender-13, Lin-11, Feng-13, Regensburger-12, Peng-14, Gao-15, Liu-16, Schindler-11, Ramezani-12, Assawaworrarit-17, Bender-13, Konotop-review}. Since the first observations of PT-symmetry breaking and power oscillations in optics \cite{El-Ganainy-07, Makris-08, Guo-09, Ruter-10}, researchers have reported rich wave phenomena unique to non-conservative systems such as unidirectionality \cite{Lin-11, Feng-13} and light transport in multifunctional devices \cite{Regensburger-12}. Related phenomena have been studied in other subfields of physics, including electrical circuits \cite{Schindler-11, Ramezani-12, Assawaworrarit-17} and mechanical oscillators \cite{Bender-13}. In the quantum regime, various aspects of PT-symmetric systems have been studied \cite{Bender-07, Graefe-08, Dast-16, Chtchelkatchev-12, YCLee-14, Chen-14, TonyLee-14, Saleur-16, Tripathi-16, Brody-16, Yin-17, Obuse-16, Obuse-17, Ashida-16, Li-16, Ashida-17}, such as Bose-Hubbard dimers \cite{Graefe-08}, entanglement \cite{Chen-14, TonyLee-14, Saleur-16}, and critical phenomena \cite{Ashida-16}. In particular, the first observation of the PT transition in quantum systems has recently been reported in ultracold atoms \cite{Li-16}.

While loss is usually considered to be detrimental to the coherence of a system \cite{Zurek-03, deVega-17, Breuer-textbook}, the unique phenomena and useful applications in PT-symmetric classical systems illustrate the utility of balanced gain and loss. From this wisdom in the classical regime, PT-symmetric quantum systems are expected to show robustness against decoherence, potentially leading to a long coherence time in quantum information processing and engineering. However, information-theoretic characterization of PT-symmetric systems has hitherto been unexplored. In addition to the practical importance, such characterization is needed for deeper understanding of PT-symmetric systems as open quantum systems.

The dynamics governed by a PT-symmetric non-Hermitian Hamiltonian $\hat{H}_{\rm PT}$ is described by \cite{Brody-12}
\begin{equation} 
\hat{\rho} \left( t \right)
= \frac{e^{- \ii \hat{H}_{\rm PT} t} \hat{\rho} \left( 0 \right) e^{\ii \hat{H}_{\rm PT}^{\dag} t}}{\mathrm{tr} \left[ e^{- \ii \hat{H}_{\rm PT} t} \hat{\rho} \left( 0 \right) e^{\ii \hat{H}_{\rm PT}^{\dag} t} \right]}.
	\label{PT-dynamics}
\end{equation}
Here we employ the usual Hilbert-Schmidt inner product and consider the effective non-unitary dynamics of an open quantum system described by $\hat{H}_{\rm PT}$ \cite{Remark-InnerProduct}. Recalling that the non-unitary state reduction enables an observer to acquire information about the system, one might well wonder if the non-unitarity of the PT dynamics given by Eq.~(\ref{PT-dynamics}) could lead to information flow between the system and the environment. In this Letter, we show that this is indeed the case, where the environment plays a role of an observer that continuously monitors the system. In particular, we demonstrate that the system in the PT-unbroken phase can completely retrieve the information that has flowed into the environment. We also find unique criticality of the information flow around the PT-transition point, across which the reversible information flow turns irreversible and vice versa. Moreover, by embedding a PT-symmetric system into a larger closed system, we identify the physical origin behind the information retrieval as an entangled partner hidden in the environment.

The information retrieval in PT-symmetric systems suggests novel possibilities for better controlling the behavior of quantum systems, in a way different from the quantum Zeno effect \cite{Zeno-1, Zeno-2, Zeno-3} or the dynamical decoupling \cite{DynamicalDecoupling-1, DynamicalDecoupling-2, DynamicalDecoupling-3}. While the dynamical decoupling relies on the time reversal by pulse injections, the information retrieval discussed in this Letter is induced by a hidden entangled partner protected by PT symmetry.  The underlying physics is essentially distinct from that of decoherence-free subspaces in which the unitary state evolution is guaranteed by certain symmetry \cite{DFS-1, DFS-2, DFS-3, DFS-4, DFS-5, DFS-6, DFS-7}; by contrast, the PT dynamics is intrinsically non-unitary both in the unbroken and broken phases.

\paragraph{Information flow.\,---}
We consider a general $N$-level quantum system and characterize information flow to and from it in terms of the trace distance between two quantum states of the same system \cite{Nielsen-Chuang}:
\begin{equation}
D \left( \hat{\rho}_{1} \left( t \right), \hat{\rho}_{2} \left( t \right) \right)
= \frac{1}{2} \mathrm{tr} \left|\,\hat{\rho}_{1} \left( t \right) - \hat{\rho}_{2} \left( t \right)\,\right|,
	\label{trace distance}
\end{equation}
where $| \hat{A} | := \sqrt{\hat{A}^{\dag} \hat{A}}$. The trace distance $D$ measures the distinguishability of two quantum states since $\left[ 1 + D \left( \hat{\rho}_{1}, \hat{\rho}_{2} \right) \right] /2$ is the maximal probability of the two states $\hat{\rho}_{1}$ and $\hat{\rho}_{2}$ being successfully distinguished \cite{distinguishability, Nielsen-05}. The trace distance is invariant under unitary transformations, i.e., no information leaks out from the system under unitary evolution. In addition, the trace distance does not increase under completely positive and trace-preserving (CPTP) maps, i.e., $D \left( \mathcal{E} \hat{\rho}_{1}, \mathcal{E} \hat{\rho}_{2} \right) \leq D \left( \hat{\rho}_{1}, \hat{\rho}_{2} \right)$ if $\mathcal{E}$ is a CPTP map \cite{CPTP}. Thus, a monotonic decrease in the distinguishability ($\dot{D} < 0$) signifies {\it unidirectional information flow} from the system to the environment. In contrast, an increase in the distinguishability ($\dot{D} > 0$) signifies {\it information backflow} from the environment to the system, implying the presence of memory effects in the open quantum dynamics. In other words, the dynamics involving a time interval with $\dot{D} > 0$ is considered to be non-Markovian \cite{Erez-08, Wolf-08, BLP-09, BLP-10, RHP-10, Luo-12, CM-14, CMM-17, BLP-16}. 

We note that the P-divisibility of linear quantum dynamical maps is equivalent to Markovianity of quantum processes \cite{RHP-10, CM-14}. Although the PT dynamics given by Eq.~(\ref{PT-dynamics}) is indeed P-divisible \cite{Brody-12}, the divisibility cannot detect the non-Markovianity of the PT dynamics due to {\it non-linearity}. Moreover, the PT dynamics in the PT-unbroken phase is not contractive. However, the time evolution of $D$ can still serve as a measure to characterize the non-Markovianity in the PT dynamics, since this measure directly quantifies information flow between a system and its environment, and hence detects the presence of memory effects even if the dynamics is neither linear nor contractive, as demonstrated below. While other measures are also possible \cite{Remark-RelativeEntropy}, the trace distance is suitable as a measure of information flow because it distinguishes all different quantum states and depends only on the system's dynamics.

\paragraph{Information retrieval.\,---}

We consider generic behavior of information flow between a PT-symmetric system and an environment. First of all, the distinguishability $D \left( t \right) := D \left( \hat{\rho}_{1} \left( t \right), \hat{\rho}_{2} \left( t \right) \right)$ oscillates and eventually returns to the initial value in the PT-unbroken phase:
\begin{equation}
\exists~T > 0~~~s.t.~~~D \left( T \right) = D \left( 0 \right).
	\label{recurrence statement}
\end{equation}
The proof of this result is given later. The recurrence of the distinguishability implies the presence of a time interval with $\dot{D} > 0$. Therefore, the system {\it retrieves the information} that has flowed into the environment, and thus the system in the PT-unbroken phase exhibits unique non-Markovian behavior. 

This non-Markovianity arises from the non-orthogonality of eigenstates. To show this, we conduct the spectral decomposition of the PT dynamics given by Eq.~(\ref{PT-dynamics}):
\begin{equation}
\hat{\rho} \left( t \right)
= \frac{\sum_{mn} \rho_{mn} e^{- \ii \left( E_{m} - E_{n} \right) t} \ket{\varphi_{m}} \bra{\varphi_{n}}}{\sum_{mn} \rho_{mn} e^{- \ii \left( E_{m} - E_{n} \right) t} \braket{\varphi_{n} | \varphi_{m}}},
	\label{spectrum decomposition}
\end{equation}
where $\ket{\varphi_{n}}$ is a right eigenstate of $\hat{H}_{\rm PT}$ with a real eigenenergy $E_{n}$ \cite{Brody-12, Brody-14}. When the dynamics is unitary, eigenstates are orthogonal to each other, and the normalization factor, which is given by the denominator of Eq.~(\ref{spectrum decomposition}), is constant. In the PT dynamics, by contrast, the normalization factor oscillates due to the non-orthogonality of eigenstates, indicating the presence of continuous information exchange between the system and the environment. In this respect, power oscillations observed in various systems \cite{Makris-08, Ruter-10, Schindler-11, Regensburger-12, Bender-13, Li-16} may be interpreted as evidence of information backflow from the environment and a signature of non-Markovianity in the PT-unbroken phase.

\paragraph{Universal criticality of information flow at the PT transition.\,---}

Unconventional criticality of information flow emerges around the PT-transition point, above which information flow ceases to be reversible. The criticality of information flow derives from the power-law behavior of an eigenenergy $E + \ii \Gamma$ of an eigenstate coalescing at the exceptional point \cite{Heiss-08}: 
\begin{equation}
E + \ii \Gamma
= E_{\rm EP} + \sum_{i=1}^{\infty} c_{i} \left( \Delta \lambda \right)^{i/p},
	\label{eigenenergy-powerlaw} 
\end{equation}
where $c_{i}$'s ($i=1,2,\cdots$) are system-specific coefficients, $\Delta \lambda$ is a real control parameter with $\Delta \lambda < 0$ ($\Delta \lambda > 0$) corresponding to the PT-unbroken (PT-broken) phase, and $p \geq 2$ is the order of the exceptional point (i.e., the number of coalescing states at the exceptional point). Note that $p$ is usually equal to $2$, but $p>2$ is also possible \cite{Heiss-12, Heiss-08, Graefe-11J, Ding-16}.

As a consequence of Eq.~(\ref{eigenenergy-powerlaw}), the recurrence time of the distinguishability given by Eq.~(\ref{recurrence statement}) exhibits the following power-law behavior near the exceptional point \cite{SuppMat}:
\begin{equation}\label{recurrence}
T \sim \left| \Delta \lambda \right|^{-1/p}~~~(\Delta \lambda \to 0^{-}).
\end{equation} 
It is remarkable that different PT-symmetric systems having the same order $p$ can exhibit the same critical behavior given by Eq.~(\ref{recurrence}). In this sense, the power-law behavior of the recurrence time is a {\it universal} property inherent in PT-symmetry breaking, where the order $p$ of the exceptional point characterizes the universality class.

In the PT-broken phase, the eigenstate with the largest imaginary part dominates the dynamics for a sufficiently long time. Thus, the distinguishability exponentially decreases and never returns to the initial value \cite{Remark-Broken-ShortTime, SuppMat}:
\begin{equation}
D \sim e^{-t/\tau}~~~\left( t \to \infty \right),
\end{equation} 
where $\tau > 0$ is a relaxation time. Thus, information flows unidirectionally from the system to the environment, and the dynamics is asymptotically Markovian. Here, the relaxation time $\tau$ also shows power-law behavior near the exceptional point:
\begin{equation}
\tau \sim \left| \Delta \lambda \right|^{-1/p}~~~( \Delta \lambda \to 0^{+} ).
\end{equation} 
The critical exponent $p$ is the same as that of the recurrence time in the PT-unbroken phase.

At the exceptional point, both the recurrence time $T$ in the PT-unbroken phase and the relaxation time $\tau$ in the PT-broken phase diverge, indicating the disappearance of a characteristic time scale of the system. As a result, the distinguishability $D$ itself exhibits power-law behavior:
\begin{equation}
D \sim t^{-\delta}~~~\left( t \to \infty \right).
\end{equation}
Here the critical exponent is $\delta=2$ if the eigenstate with the second largest imaginary part in the PT-broken phase coalesces with some other states at the exceptional point; otherwise, $\delta = p-1$ \cite{SuppMat}.

The above criticality is universal in the sense that it depends only on the order of the exceptional point. The obtained universal behavior accompanying diverging characteristic scales of the system is reminiscent of the critical phenomena in many-body systems \cite{Goldenfeld, Sachdev}. It is a unique feature of non-Hermitian systems that such critical behavior can appear even in finite systems, in sharp contrast to Hermitian systems where critical behavior emerges only in the thermodynamic limit.

\paragraph{Embedding a PT-symmetric system into a larger Hermitian system.\,---}

To establish clear physical understanding of the information retrieval, we consider embedding a PT-symmetric system of interest into a larger Hermitian system. Here, our idea is motivated by the Naimark extension for quantum measurement \cite{Naimark, Ozawa, Hayashi}: by adding an ancilla (a measuring apparatus) and extending the Hilbert space, any non-unitary dynamics can be understood as a unitary dynamics of the entire system followed by quantum measurement acting on the ancilla. Moreover, the extension provides a rigorous proof of the information retrieval given by Eq.~(\ref{recurrence statement}).

To begin with, we consider a case of the PT-unbroken phase. We examine pure states for the sake of simplicity, but the discussion can straightforwardly be generalized to mixed states. We introduce a Hermitian invertible operator $\hat{\eta}$ that characterizes the pseudo-Hermiticity, satisfying $\hat{\eta} \hat{H}_{\rm PT} = \hat{H}_{\rm PT}^{\dag} \hat{\eta}$ \cite{Most-02-a, Most-02-b}. Note that the unbrokenness of PT symmetry is equivalent to the existence of an invertible linear operator $\hat{O}$ such that $\hat{\eta} = \hat{O}\hat{O}^{\dag}$ \cite{Most-02-b}. We define a parameter $c := \sum_{i=1}^{N} 1/\lambda_{i}$ and an operator $\hat{\zeta} := c\hat{\eta} - \hat{I}_{N}$, where $\lambda_{i}$'s ($i=1,2,\cdots,N$) are eigenvalues of $\hat{\eta}$. If PT symmetry is unbroken, $\hat{\eta}$ is positive, so is $\hat{\zeta}$. An important property that allows for the following extension is that $\braket{\psi_{\rm PT} | \hat{\eta} | \psi_{\rm PT}}$ is invariant under the PT dynamics $\ket{\psi_{\rm PT} \left( t \right)} := e^{- \ii \hat{H}_{\rm PT} t} \ket{\psi_{0}}$ \cite{Most-02-a}.

With the operator $\hat{\zeta}$ defined above, we construct a {\it PT-symmetric counterpart} $\hat{\zeta}^{1/2} \ket{\psi_{\rm PT} \left( t \right)}$, add a two-level ancilla, and consider an entangled state $\ket{{\sf \Psi}_{\rm tot} \left( t \right)}$ in an extended Hilbert space $\mathcal{H}_{2} \otimes \mathcal{H}_{\rm PT}$:
\begin{equation}
\ket{{\sf \Psi}_{\rm tot} \left( t \right)}
= \ket{\uparrow} \otimes \ket{\psi_{\rm PT} \left( t \right)} + \ket{\downarrow} \otimes \left( \hat{\zeta}^{1/2} \ket{\psi_{\rm PT} \left( t \right)} \right).
\end{equation}
The norm of such a state in the extended space is $\braket{{\sf \Psi_{\rm tot}} | {\sf \Psi_{\rm tot}}} = c \braket{\psi_{\rm PT} | \hat{\eta} | \psi_{\rm PT}}$ and is conserved for an arbitrary PT dynamics. Therefore, the entire extended system is Hermitian and closed. When a measurement is performed on the ancilla and $\ket{\uparrow}$ is postselected, we obtain $\ket{\psi_{\rm PT} \left( t \right)}$; when $\ket{\downarrow}$ is postselected, we obtain the PT-symmetric counterpart $\hat{\zeta}^{1/2} \ket{\psi_{\rm PT} \left( t \right)}$. While $\ket{\psi_{\rm PT} \left( t \right)}$ obeys $\hat{H}_{\rm PT}$, its counterpart $\hat{\zeta}^{1/2} \ket{\psi_{\rm PT} \left( t \right)}$ obeys $\hat{\zeta}^{1/2} \hat{H}_{\rm PT} \hat{\zeta}^{-1/2}$. Our extension is minimal since the ancilla as an environment has only two levels. In particular, if $\hat{H}_{\rm PT}$ is a two-level system, $\hat{\zeta}^{1/2}$ reduces to $\hat{\eta}$ and $\hat{\zeta}^{1/2} \hat{H}_{\rm PT} \hat{\zeta}^{-1/2} = \hat{H}_{\rm PT}^{\dag}$ \cite{GS-08}. An experiment using such an extension for a two-level system has recently been realized with two entangled photons \cite{Tang-16}. We emphasize that our idea of the extension is different from that of Ref. \cite{GS-08}, which essentially relies on the property $\hat{\eta} + \hat{\eta}^{-1} = \left( {\rm tr}\,\hat{\eta} \right) \hat{I}_{2}$ unique to a two-level system. Moreover, we consider the extension in the PT-broken phase as shown below, while Ref. \cite{GS-08} is restricted to the PT-unbroken phase.

The conservation of the norm of the extended state $\ket{{\sf \Psi}_{\rm tot} \left( t \right)}$ implies the unitarity of its evolution ${\sf U}_{\rm tot}$. We can explicitly construct the Hermitian Hamiltonian ${\sf H}_{\rm tot}$ in the extended closed system that satisfies ${\sf U}_{\rm tot} = e^{-\ii {\sf \hat{H}}_{\rm tot} t}$:
$
{\sf \hat{H}}_{\rm tot}
= \hat{I}_{2} \otimes \hat{H}_{S} + \hat{H}_{I},
$
where $\hat{H}_{S}$ acts locally on the Hilbert space of the system $\mathcal{H}_{\rm PT}$, and $\hat{H}_{I}$ acts globally on $\mathcal{H}_{2} \otimes \mathcal{H}_{\rm PT}$ as an interaction between the system and the ancilla \cite{SuppMat}. A crucial property of the extended Hamiltonian is that the original system $\hat{H}_{\rm PT}$ is non-Hermitian if and only if the characteristic interaction $\hat{H}_{I}$ in the extended Hilbert space is non-zero. Since the interaction is global, the entanglement (or quantum mutual information) between the system and the ancilla oscillates in time. The presence of an {\it entangled partner hidden in the environment} is the physical origin of the memory effect of PT-symmetric systems: the information that has flowed into the environment is actually stored in the entanglement with the ancilla.

\paragraph{PT-symmetry breaking viewed from a larger closed system.\,---}

The information retrieval can now be proven. The extended Hermitian system with finite energy levels returns back to the original state by the quantum recurrence theorem \cite{Recurrence-1, Recurrence-2}, and the PT-symmetric system obtained by the projection of an extended Hermitian system also returns to the initial state. Here, we emphasize that it is {\it a priori} not obvious whether such an extension with a finite-level ancilla is possible, and this underlying finite dimensionality is a nontrivial universal feature of the PT-unbroken open quantum systems. When PT symmetry is broken, in contrast, $\hat{\eta}$ is no longer Hermitian and the above extension breaks down. In fact, the irreversibility of information flow in the PT-broken phase leads to the impossibility of the extension with a finite-dimensional ancilla, and an {\it infinite-dimensional} ancilla is needed for embedding a PT-broken system into a larger closed system. The recurrence theorem cannot be applied to an infinite system, where the irreversible dynamics should emerge \cite{Remark-Weisskopf-Wigner}.

The above extension of a PT-symmetric system gives an insight into PT-symmetry breaking. PT-symmetry breaking can occur in finite systems \cite{Bender-review}, although the conventional theory of phase transitions tells us that spontaneous symmetry breaking occurs only in the thermodynamic (infinite-dimensional) limit \cite{Goldenfeld, Sachdev}. The above observation implies a close relationship of PT-symmetry breaking to this conventional wisdom: in view of an extended Hermitian system, an {\it infinite-dimensional} ancilla is required for the PT transition to occur. The exceptional point is characterized as a singular point at which the dimension of the companion ancilla suddenly changes from finite to infinite. 

It is also intriguing to revisit the criticality of information flow at the exceptional point in an extended Hermitian system. At the exceptional point, the norm of the PT-symmetric counterpart diverges and the probability of successful postselection of the system of interest vanishes. Peculiar as it may appear, the vanishing probability at the exceptional point is reminiscent of the conventional critical phenomena. Around the critical point, the system undergoes larger fluctuations and hence the number of measurements required for obtaining the true value increases critically according to a power law.

\paragraph{Two-level system.\,---}

\begin{figure}[t]
\includegraphics[width=86mm]{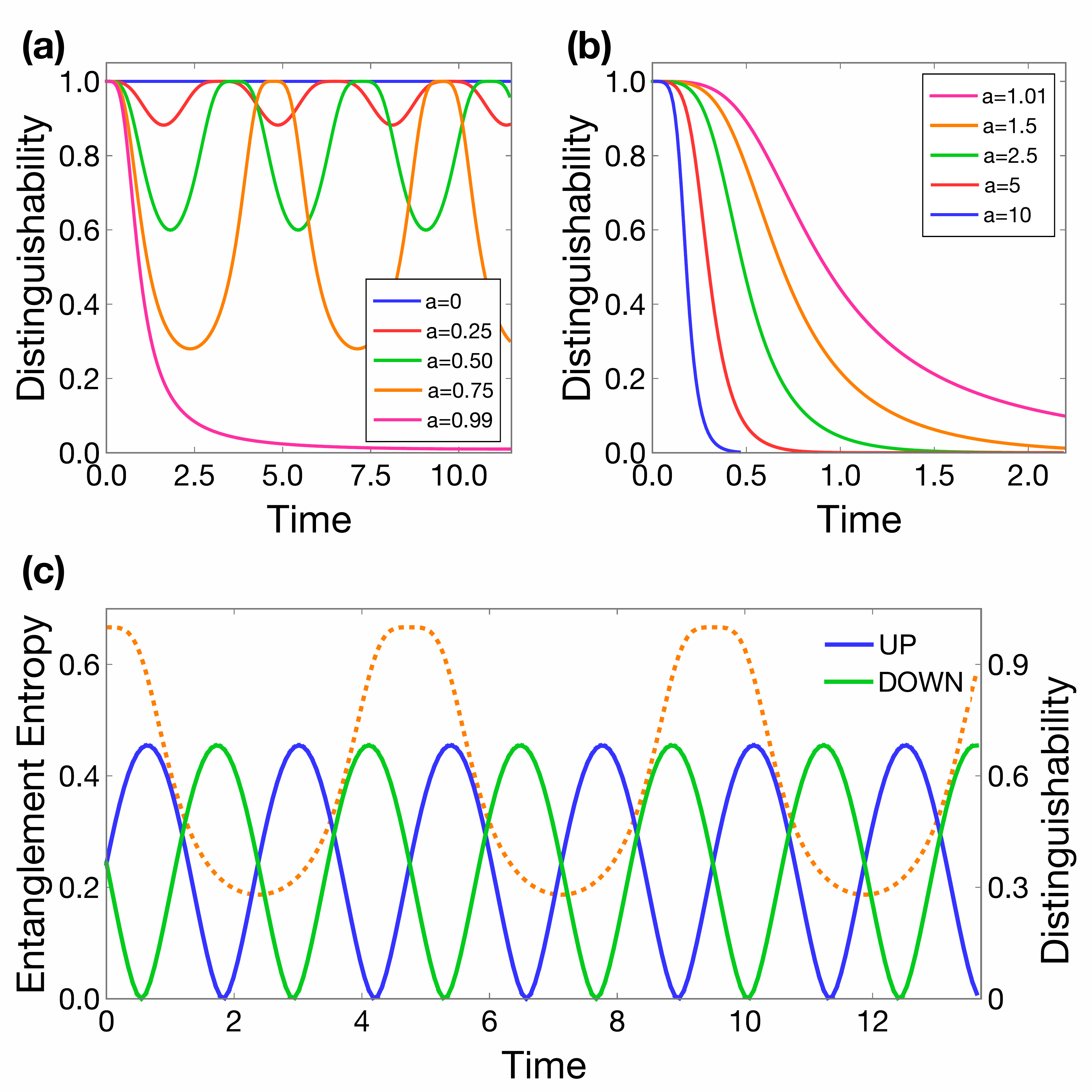} 
\caption{Information flow in the PT-symmetric two-level system. (a) Time evolution of the distinguishability $D \left( t \right)$ in the PT-unbroken phase ($0 < a < 1$). The distinguishability oscillates with period $T = \pi/\sqrt{1-a^{2}}$. The more closely we approach the exceptional point ($a=1$), the longer the period becomes and the larger the degree of the information retrieval grows. (b) Time evolution of the distinguishability in the PT-broken phase ($a > 1$).  The distinguishability decays with a relaxation time $\tau = 1/ \left( 2\sqrt{a^{2}-1} \right)$. Near the exceptional point, $D \left( t \right)$ asymptotically behaves as $D \sim 1/t^{2}$. (c) Time evolution of the entanglement entropy between the system and the ancilla with $a=0.75$. The blue (green) solid curve represents the entanglement entropy for the PT dynamics whose initial state is $\ket{\uparrow}$ ($\ket{\downarrow}$), while the dotted curve represents the distinguishability between them. The entanglement clearly oscillates with period $T_{E} = \pi / \left( 2 \sqrt{1-a^{2}} \right)$, which is one half of the period of $D \left( t \right)$.}
	\label{fig: two-level}
\end{figure}

As the simplest example, we consider a two-level system $\hat{H}_{\rm PT}
= s \left( \hat{\sigma}_{x} + \ii a\,\hat{\sigma}_{z} \right)$, where $s \geq 0$ is an energy scale and $a \geq 0$ represents the degree of non-Hermiticity. This model has been realized in both classical \cite{Ruter-10, Gao-15, Liu-16} and quantum \cite{Tang-16, Li-16} experiments. The eigenenergies are $\pm s \sqrt{1-a^{2}}$, and thus the exceptional point is located at $a=1$ with the order $p=2$. Let $\ket{\uparrow \left( t \right)}$ ($\ket{\downarrow \left( t \right)}$) be the time-evolved state starting from the initial state $\ket{\uparrow}$ ($\ket{\downarrow}$). Then the distinguishability between them is calculated to be \cite{SuppMat}
\begin{equation}
D \left( t \right)
= \left[ 1 + \left( \frac{2a \sin^{2} \left( \sqrt{1-a^{2}} st \right)}{1-a^{2}} \right)^{2} \right]^{-1/2}.
\end{equation}
We thus confirm that the obtained behavior is consistent with the general theory discussed above (Fig.~\ref{fig: two-level}\,(a)\,(b)).

The presence of a hidden entangled partner behind the information retrieval is understood through examination of the entanglement entropy between the system and the ancilla. Figure \ref{fig: two-level}\,(c) represents the calculated entanglement entropy, which clearly shows the oscillation of the entanglement synchronous with that of $D \left( t \right)$.

\paragraph{PT-symmetric optics.\,---}

The above discussion on information flow can be applied to classical optical systems \cite{El-Ganainy-07, Makris-08}. Let us consider the propagation of beams through a complex PT-symmetric potential $V_{\rm PT} \left( x \right)$, where their electric-field envelopes obey the paraxial equation of diffraction $\ii \, \partial_{z} \psi \left( x, z \right) = - \left[ \partial_{x}^{2} + V_{\rm PT} \left( x \right) \right] \psi \left( x, z \right)$. Then the distinguishability between two electric-field envelopes of the same system corresponds to
$
D \left( z \right)
= \sqrt{1 - \frac{\left| \left< \psi, \phi \right> \right|^{2}}{\left< \psi, \psi \right> \left< \phi, \phi \right>}},
$
where the inner product $\left< \cdot, \cdot \right>$ is defined by $\left< \psi, \phi \right> \left( z \right) := \int dx~\psi^{*} \left( x, z\right) \phi \left( x, z\right)$.

The earlier discussions cannot directly be applied to optical settings in a strict sense, since the discussions are confined to finite-dimensional systems. However, we expect that the universal behavior of information flow continues to hold, and actually confirm it for a specific PT-symmetric periodic potential $V_{\rm PT} \left( x \right) = V_{0} \left[ \cos \left( 2\pi x/a \right) + \ii\,\lambda \sin \left( 2 \pi x/a\right) \right]$, where $\lambda \geq 0$ measures the degree of non-Hermiticity \cite{Makris-08}. In the PT-unbroken phase ($0 < \lambda < 1$), the distinguishability oscillates in accordance with power oscillations \cite{Makris-08, Ruter-10, Schindler-11, Regensburger-12, Bender-13}; in the PT-broken phase ($\lambda > 1$), it exponentially decays because of attenuation and amplification. A remarkable feature shows up at the exceptional point ($\lambda = 1$) \cite{Longhi-10, Graefe-11}, where the distinguishability asymptotically decays according to a power law whose critical exponent is $1$ or $2$ for appropriately chosen initial states \cite{SuppMat}.

\paragraph{Conclusion and outlook.\,---}

We have found that information that has flowed into an environment can be retrieved in a PT-symmetric system. This information retrieval originates from a universal feature of an environment that couples to a PT-symmetric system: it can be modeled by a finite-dimensional (infinite-dimensional) ancilla in the PT-unbroken (PT-broken) phase. Here the exceptional point plays a role of the critical point around which many physical quantities such as the recurrence time and the distinguishability show power-law behavior. These findings may find novel applications to quantum control. 

It is also worthwhile to remark that the information retrieval in a PT-symmetric system might be distinct from the conventional non-Markovian behavior that has been found in typical open quantum systems coupled to an infinite bath \cite{Zurek-03, deVega-17, Breuer-textbook, Erez-08, Wolf-08, BLP-09, BLP-10, RHP-10, Luo-12, CM-14, CMM-17, BLP-16}. While we expect that a broad class of information measures can be used as a witness of the non-Markovianity discussed in this Letter, it merits further study to clarify their applicability in detail and explore possible relevance to the known non-Markovian behavior mentioned above.

\paragraph{Acknowledgements.\,---}

We thank Y\^uto Murashita and Sho Higashikawa for valuable discussions. This work was supported by KAKENHI Grant No. JP26287088 from the Japan Society for the Promotion of Science, a Grant-in-Aid for Scientific Research on Innovative Areas ``Topological Materials Science" (KAKENHI Grant No. JP15H05855), and the Photon Frontier Network Program from MEXT of Japan. K. K. and Y. A. were supported by the JSPS through Program for Leading Graduate Schools (ALPS). Y. A. acknowledges support from JSPS (Grant No. JP16J03613).


\widetext
\pagebreak
\begin{center}
\textbf{\large Supplemental Material}
\end{center}

\renewcommand{\theequation}{S\arabic{equation}}
\renewcommand{\thefigure}{S\arabic{figure}}
\renewcommand{\bibnumfmt}[1]{[S#1]}
\setcounter{equation}{0}
\setcounter{figure}{0}

\bigskip
\begin{center}
\textbf{Proof of Eq.~(6)}
\end{center}

We derive Eq.~(6) in the main text which shows the power-law behavior of the recurrence time in the PT-unbroken phase. From the spectral decomposition of the PT dynamics
\begin{equation}
\hat{\rho} \left( t \right)
= \frac{\sum_{mn} \rho_{mn} e^{- \ii \omega_{mn} t} \ket{\varphi_{m}} \bra{\varphi_{n}}}{\sum_{mn} \rho_{mn} e^{- \ii \omega_{mn} t} \braket{\varphi_{n} | \varphi_{m}}}~~~\left( \omega_{mn} := E_{m} - E_{n} \right),
\end{equation}
the recurrence time is evaluated as the least common multiple of $2\pi/\omega_{mn}$ ($m,\,n = 1,\,2,\,\cdots,\,N$). Note that the vanishing energy difference $\Delta \omega$ between coalescing $p$ states behaves near the exceptional point as \cite{Heiss-08}
\begin{equation}
\Delta \omega \sim \left| \Delta \lambda \right|^{1/p}~~~\left( \Delta \lambda \to 0^{-} \right).
\end{equation}
Thus, the recurrence time $T \sim 2\pi/\Delta\omega$ diverges at the exceptional point as in Eq.~(6). $\blacksquare$

\bigskip
\begin{center}
\textbf{Proof of Eqs.~(7) and (8)}
\end{center}

We derive Eqs.~(7) and (8) in the main text which show the exponential decay of the distinguishability and the power-law behavior of the relaxation time in the PT-broken phase. We set a right (left) eigenstate of $\hat{H}_{\rm PT}$ with an eigenenergy $E_{n} + \ii \Gamma_{n}$ as $\ket{\varphi_{n}}$ ($\ket{\chi_{n}}$) \cite{Brody-14}:
\begin{equation}
\hat{H}_{\rm PT} \ket{\varphi_{n}} = \left( E_{n} + \ii \Gamma_{n} \right) \ket{\varphi_{n}},~~
\bra{\chi_{n}} \hat{H}_{\rm PT} = \left( E_{n} + \ii \Gamma_{n} \right) \bra{\chi_{n}}.
\end{equation}
The eigenstates are normalized with $\braket{\varphi_{n} | \varphi_{n}} = \braket{\chi_{n} | \chi_{n}} = 1$. We define the right (left) eigenstate with the largest imaginary part in the PT-broken phase as $\ket{\varphi_{1}}$ ($\ket{\chi_{1}}$), and the right (left) eigenstate with the second largest imaginary part as $\ket{\varphi_{2}}$ ($\ket{\chi_{2}}$). For the eigenstate coalescing at the exceptional point, the imaginary part of the eigenenergy can be expanded as
\begin{equation}
\Gamma = \sum_{i=1}^{\infty} {a}_{i} \left( \Delta \lambda \right)^{i/p},
	\label{EP imaginary power-law}
\end{equation}
where $a_{i}$ ($i = 1,\,2,\,\cdots$) are system-specific coefficients \cite{Heiss-08}. In addition, the eigenstate can also be expanded as
\begin{equation}
\ket{\varphi_{n}} = \ket{\rm EP} + \sum_{i=1}^{\infty} \left( \Delta \lambda \right)^{i/p} \ket{\psi_{n}^{\left( i \right)}},~~
\ket{\chi_{n}} = \ket{\rm \overline{EP}} + \sum_{i=1}^{\infty} \left( \Delta \lambda \right)^{i/p} \ket{\overline{\psi}_{n}^{\left( i \right)}},
		\label{EP state power-law}
\end{equation}
where $\ket{\rm EP}$ ($\ket{\rm \overline{EP}}$) is the right (left) eigenstate at the exceptional point normalized with $\braket{\rm EP | EP} = \braket{\rm \overline{EP} | \overline{EP}} = 1$, and $\ket{\psi_{n}^{\left( i \right)}}$ and $\ket{\overline{\psi}_{n}^{\left( i \right)}}$ ($i=1,2,\cdots$) are unnormalized vectors. What is crucial for the criticality at the exceptional point lies in the power-law behavior of the eigenenergies and eigenstates near the exceptional point. 

Omitting the normalization factor, the PT dynamics is represented as
\begin{equation}
\hat{\rho} \left( t \right)
\propto \sum_{ij} \rho_{ij} e^{-\ii \left( E_{i} - E_{j} \right) t} e^{\left( \Gamma_{i} + \Gamma_{j} \right) t} \ket{\varphi_{i}} \bra{\varphi_{j}},~~
\rho_{ij} := \frac{\braket{\chi_{i} |\,\hat{\rho}_{0}\,| \chi_{j}}}{\braket{\chi_{i} | \varphi_{i}} \braket{\varphi_{j} | \chi_{j}}}.
\end{equation} 
After a sufficiently long time, both $\ket{\varphi_{1}}$ and $\ket{\varphi_{2}}$ dominate the dynamics, and hence the PT dynamics asymptotically behaves as 
\begin{equation}
\hat{\rho} \left( t \right)
\propto \ket{\varphi_{1}} \bra{\varphi_{1}} + \left[ C_{\rho} e^{-\ii \omega t} \ket{\varphi_{1}} \bra{\varphi_{2}} + {\rm H.c.} \right] e^{- \Delta \Gamma \cdot t} + o \left( e^{- \Delta \Gamma \cdot t} \right),
\end{equation}
where $C_{\rho} := \rho_{12}/\rho_{11}$ and $\omega := E_{1} - E_{2}$. Including the normalization factor, we have
\begin{eqnarray}
\hat{\rho} \left( t \right)
&\sim& \frac{\ket{\varphi_{1}} \bra{\varphi_{1}} + \left[ C_{\rho} e^{-\ii \omega t} \ket{\varphi_{1}} \bra{\varphi_{2}} + {\rm H.c.} \right] e^{- \Delta \Gamma \cdot t}}{{\rm tr} \left[ \ket{\varphi_{1}} \bra{\varphi_{1}} + \left[ C_{\rho} e^{-\ii \omega t} \ket{\varphi_{1}} \bra{\varphi_{2}} + {\rm H.c.} \right] e^{- \Delta \Gamma \cdot t} \right]} \nonumber \\
&=& \frac{\ket{\varphi_{1}} \bra{\varphi_{1}} + \left[ C_{\rho} e^{-\ii \omega t} \ket{\varphi_{1}} \bra{\varphi_{2}} + {\rm H.c.} \right] e^{- \Delta \Gamma \cdot t}}{1 + \left[ C_{\rho} e^{-\ii \omega t} \braket{\varphi_{2} | \varphi_{1}} + {\rm c.c.} \right] e^{- \Delta \Gamma \cdot t}} \nonumber \\
&\simeq& \ket{\varphi_{1}} \bra{\varphi_{1}} + \left\{ \left[ C_{\rho} e^{-\ii \omega t} \ket{\varphi_{1}} \bra{\varphi_{2}} + {\rm H.c.} \right] - \left[ C_{\rho} e^{-\ii \omega t} \braket{\varphi_{2} | \varphi_{1}} + {\rm c.c.} \right] \ket{\varphi_{1}} \bra{\varphi_{1}} \right\} e^{- \Delta \Gamma \cdot t}.
\end{eqnarray}
We then define a normalized state $\ket{\varphi_{\times}}$ that is a linear superposition of $\ket{\phi_{1}}$ and $\ket{\phi_{2}}$ and orthogonal to the state $\ket{\varphi_{1}}$:
\begin{equation}
\ket{\varphi_{2}}
= \braket{\varphi_{1} | \varphi_{2}} \ket{\varphi_{1}}
+ \sqrt{1 - \left| \braket{\varphi_{1} | \varphi_{2}} \right|^{2}} \ket{\varphi_{\times}}.
\end{equation}
Using $\ket{\varphi_{\times}}$, $\hat{\rho} \left( t \right)$ becomes after a sufficiently long time
\begin{equation}
\hat{\rho} \left( t \right)
\sim \ket{\varphi_{1}} \bra{\varphi_{1}}
+ \sqrt{1 - \left| \braket{\varphi_{1} | \varphi_{2}} \right|^{2}} \left[ C_{\rho} e^{-\ii \omega t} \ket{\varphi_{1}} \bra{\varphi_{\times}} + {\rm H.c.} \right] e^{- \Delta \Gamma \cdot t}.
\end{equation}
Therefore, the trace distance between two arbitrary quantum states $\hat{\rho} \left( t \right)$ and $\hat{\sigma} \left( t \right)$ in the PT-broken phase asymptotically behaves as follows:
\begin{eqnarray}
D \left( t \right)
&=& \frac{1}{2}\,\mathrm{tr} \left| \hat{\rho} \left( t \right) - \hat{\sigma} \left( t \right) \right| \nonumber \\
&\sim& \frac{1}{2} \sqrt{1 - \left| \braket{\varphi_{1} | \varphi_{2}} \right|^{2}}\,{\rm tr} 
\left| \left( C_{\rho} - C_{\sigma} \right) e^{-\ii \omega t} \ket{\varphi_{1}} \bra{\varphi_{\times}} + {\rm H.c.} \right| e^{- \Delta \Gamma \cdot t} \nonumber \\
&=& \left| C_{\rho} - C_{\sigma} \right| \sqrt{1 - \left| \braket{\varphi_{1} | \varphi_{2}} \right|^{2}} e^{- \Delta \Gamma \cdot t}.
\end{eqnarray}
Finally, we obtain $\Delta \Gamma \sim \left( \Delta \lambda \right)^{1/p}$ near the exceptional point due to Eq.~(\ref{EP imaginary power-law}), which completes the proof. $\blacksquare$

\bigskip
\begin{center}
\textbf{Proof of Eq.~(9)}
\end{center}

We derive Eq.~(9) in the main text which shows the power-law decay of the distinguishability at the exceptional point. We first show the following Lemma.

\begin{figure}[H]
\centering
\includegraphics[width=120mm]{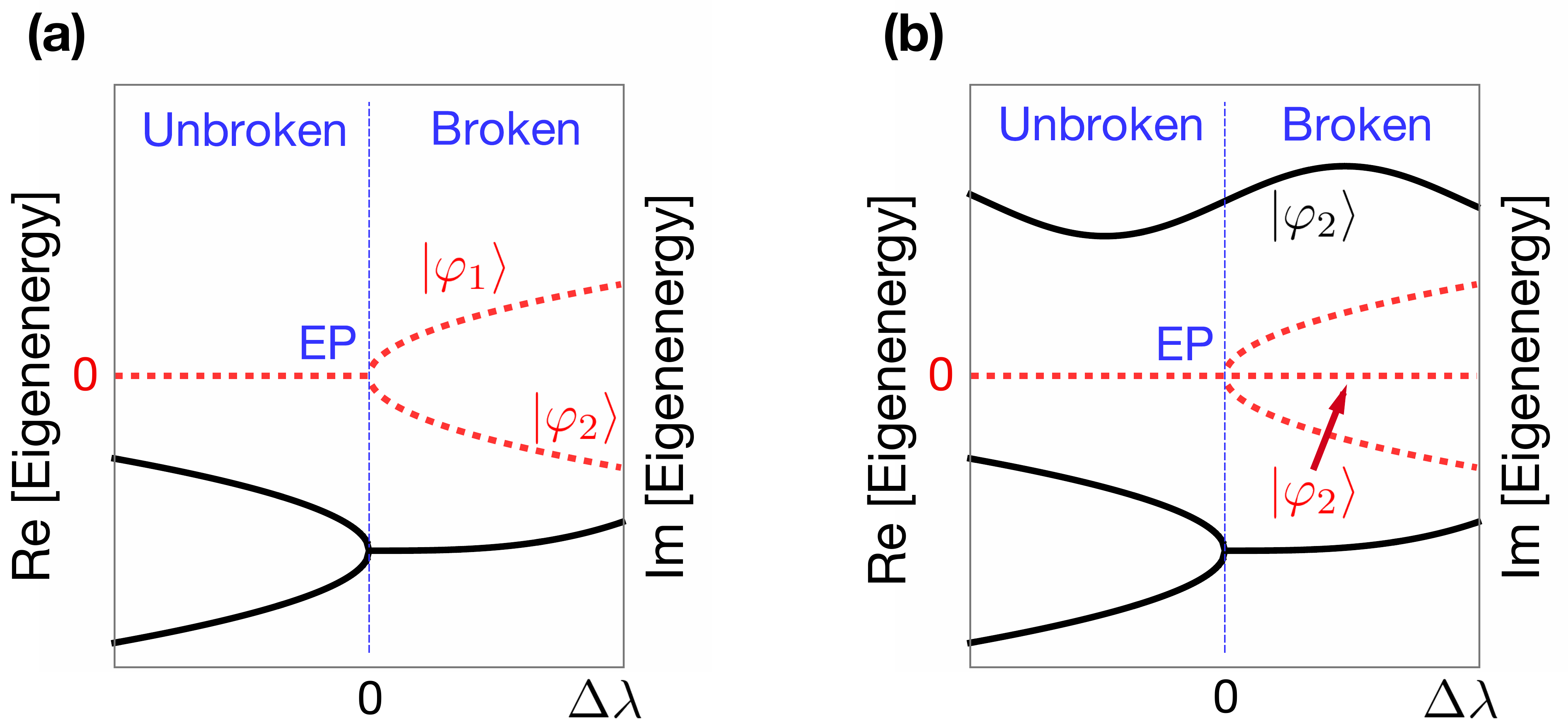} 
\caption{Exceptional points (EPs) are classified into two cases by the eigenstate with the second largest imaginary part of its eigenvalue in the PT-broken phase (labeled as $\ket{\varphi_{2}}$). (a) The state $\ket{\varphi_{2}}$ and some other states (here labeled $\ket{\varphi_{1}}$) coalesce at the exceptional point. (b) The state $\ket{\varphi_{2}}$ stands alone and does not coalesce with other states at the exceptional point as can be seen from the real part of its eigenvalue.}
	\label{fig: EP}
\end{figure}

\bigskip
\noindent{\bf Lemma}~~~The function $f \left( \Delta \lambda \right) := \left| C_{\rho} - C_{\sigma} \right| \sqrt{1 - \left| \braket{\varphi_{1} | \varphi_{2}} \right|^{2}}$ given in the proof of Eqs.~(7) and (8) exhibits the following power-law behavior near the exceptional point:
\begin{equation}
f \left( \Delta \lambda \right) \sim \left| \Delta \lambda \right|^{\delta/p}~~~\left( \Delta \lambda \to 0^{+} \right),
\end{equation}
where $\delta=2$ if $\ket{\varphi_{2}}$ coalesces at the exceptional point (Fig.~\ref{fig: EP}\,(a)), and $\delta =p-1$ otherwise (Fig.~\ref{fig: EP}\,(b)).

\noindent\textbf{Proof of Lemma}~~~Around the exceptional point, both $\left| C_{\rho} - C_{\sigma} \right|$ and  $\sqrt{1 - \left| \braket{\varphi_{1} | \varphi_{2}} \right|^{2}}$ in the asymptotic formula given in the proof of Eqs.~(7) and (8) can exhibit power-law behavior. We first focus on $\left| C_{\rho} - C_{\sigma} \right|$ and prove its criticality
\begin{equation}
\left| C_{\rho} - C_{\sigma} \right|
\sim \left( \Delta \lambda \right)^{e_{1}/p}~~~\left( \Delta \lambda \to 0^{+} \right),
\end{equation} 
where $e_{1} = 1$ if $\ket{\varphi_{2}}$ coalesces at the exceptional point, and $e_{1} =p-1$ otherwise. First we note that
\begin{equation}
\left| C_{\rho} - C_{\sigma} \right|
= \left( \frac{\braket{\chi_{1} |\,\hat{\rho}_{0}\,| \chi_{2}}}{\braket{\chi_{1} |\,\hat{\rho}_{0}\,| \chi_{1}}}
- \frac{\braket{\chi_{1} |\,\hat{\sigma}_{0}\,| \chi_{2}}}{\braket{\chi_{1} |\,\hat{\sigma}_{0}\,| \chi_{1}}} \right) \frac{\braket{\varphi_{1} | \chi_{1}}}{\braket{\varphi_{2} | \chi_{2}}}.
\end{equation}
On the one hand, when $\ket{\varphi_{2}}$ coalesces and converges to $\ket{\rm EP}$ at the exceptional point, we have
\begin{equation}
\frac{\braket{\chi_{1} |\,\hat{\rho}_{0}\,| \chi_{2} } }{\braket{\chi_{1} |\,\hat{\rho}_{0}\,| \chi_{1}}}
\simeq 1 + \frac{ \bra{ \overline{\mathrm{EP}} } \hat{\rho}_{0} \left( \ket{ \overline{\psi}_{2}^{\left( 1 \right)}} - \ket{ \overline{\psi}_{1}^{\left( 1 \right)}} \right) }{\braket{ \overline{\mathrm{EP}} |\,\hat{\rho}_{0}\,| \overline{\mathrm{EP}} }} \lambda^{1/p},~~
\frac{\braket{\varphi_{1} | \chi_{1}}}{\braket{\varphi_{2} | \chi_{2}}} \sim 1,
\end{equation}
and thus we obtain the critical exponent $e_{1}=1$. On the other hand, when $\ket{\varphi_{2}}$ converges to $\ket{n} \neq \ket{\rm EP}$ at the exceptional point, $\braket{\varphi_{1} | \chi_{1}} \sim \lambda^{1-1/p}$ and
\begin{equation}
\left| C_{\rho} - C_{\sigma} \right|
\sim \left( \frac{\braket{ \overline{\mathrm{EP}} |\,\hat{\rho}_{0}\,| n}}{\braket{ \overline{\mathrm{EP}} |\,\hat{\rho}_{0}\,| \overline{\mathrm{EP}}}}
- \frac{\braket{ \overline{\mathrm{EP}} |\,\hat{\sigma}_{0}\,| n}}{\braket{\overline{\mathrm{EP}} |\,\hat{\sigma}_{0}\,| \overline{\mathrm{EP}}}} \right) \frac{\lambda^{1-1/p}}{\braket{n | \tilde{n}}}
\sim \lambda^{1-1/p}.
\end{equation}
This indicates that the critical exponent is $e_{1} = p-1$.

\smallskip
We now show the critical behavior of $\sqrt{1 - \left| \braket{\varphi_{1} | \varphi_{2}} \right|^{2}}$: 
\begin{equation}
\sqrt{1 - \left| \braket{\varphi_{1} | \varphi_{2}} \right|^{2}}
\sim \left( \Delta \lambda \right)^{e_{2}/p}~~~\left( \Delta \lambda \to 0^{+} \right),
\end{equation} 
where $e_{2} = 1$ if $\ket{\varphi_{2}}$ coalesces at the exceptional point, and $e_{2} = 0$ otherwise. On the one hand, when $\ket{\varphi_{2}}$ converges to $\ket{\rm EP}$ at the exceptional point, we have the following formula for $\braket{\varphi_{1} | \varphi_{2}}$\,:
\begin{eqnarray}
\braket{\varphi_{1} | \varphi_{2}}
\sim 1 
+ \left( \Delta \lambda \right)^{1/p} \left( \braket{ \psi_{1}^{\left( 1 \right)} | \mathrm{EP}} + \braket{ \mathrm{EP} | \psi_{2}^{\left( 1 \right)}} \right) 
+ \left( \Delta \lambda \right)^{2/p} \left( \braket{\psi_{1}^{\left( 1 \right)} | \psi_{2}^{\left( 1 \right)}} + \braket{ \psi_{1}^{\left( 2 \right)} | \mathrm{EP}} + \braket{ \mathrm{EP} | \psi_{2}^{\left( 2 \right)}} \right).
	\label{expansion}
\end{eqnarray}
Noting the expansion in Eq.~(\ref{EP state power-law}) for the coalescing eigenstates, we obtain
\begin{equation}
\braket{\varphi_{n} | \varphi_{n}} \simeq \braket{\mathrm{EP} | \mathrm{EP}}
+ \left( \Delta \lambda \right)^{1/p} \left( \braket{ \psi_{n}^{\left( 1 \right)} | \mathrm{EP}} + \braket{ \mathrm{EP} | \psi_{n}^{\left( 1 \right)}} \right)
+ \left( \Delta \lambda \right)^{2/p} \left( \braket{\psi_{n}^{\left( 1 \right)} | \psi_{n}^{\left( 1 \right)}} + \braket{ \psi_{n}^{\left( 2 \right)} | \mathrm{EP}} + \braket{ \mathrm{EP} | \psi_{n}^{\left( 2 \right)}} \right).
\end{equation}
Due to the normalization condition $\braket{\varphi_{n} | \varphi_{n}} = \braket{\mathrm{EP} | \mathrm{EP}} = 1$, the comparison of the coefficients in the expansion of $\left( \Delta \lambda \right)^{1/p}$ leads to
\begin{equation}
\mathrm{Re} \left[ \braket{ \psi_{n}^{\left( 1 \right)} | \mathrm{EP}} \right] = 0,~~
\braket{\psi_{n}^{\left( 1 \right)} | \psi_{n}^{\left( 1 \right)}} + 2\,\mathrm{Re} \left[ \braket{ \psi_{n}^{\left( 2 \right)} | \mathrm{EP}} \right]
 = 0.
 	\label{Eq abc}
\end{equation}
Therefore, in the expansion of $\left( \Delta \lambda \right)^{1/p}$ in Eq.~(\ref{expansion}), the zeroth-order term of $\left| \braket{\varphi_{1} | \varphi_{2}}\right|^{2}$ is $1$, the first-order term is $0$ due to Eq.~(\ref{Eq abc}) for $n=1,2$, and the second-order term is non-zero in general. Hence, the critical exponent is determined to be $e_{2} = 1$. On the other hand, when $\ket{\varphi_{2}}$ converges to $\ket{n} \neq \ket{\rm EP}$ at the exceptional point, $\braket{\varphi_{1} | \varphi_{2}} \sim \braket{{\rm EP}|n}$ and thus the zeroth-order term of $\left| \braket{\varphi_{1} | \varphi_{2}}\right|^{2}$ is different from $1$ in general. This indicates that the critical exponent is $e_{2} = 0$. $\blacksquare$

\bigskip
The distinguishability can be represented as
\begin{equation}
D \left( t \right)
= f \left( \Delta \lambda \right) g \left( \frac{t}{\tau} \right),
\end{equation}
where $g$ is a function determined by the detail of the system. Around the exceptional point, combining Eq.~(8) in the main text and Lemma leads to the asymptotic behavior of the distinguishability
\begin{equation}
D \left( t \right)
\sim \left| \Delta \lambda \right|^{\delta/p} \cdot g \left( \frac{t}{\left| \Delta \lambda \right|^{-1/p}} \right) \sim t^{-\delta}.~\blacksquare
\end{equation}

\newpage
\begin{center}
\textbf{Explicit form of the extended Hermitian Hamiltonian}
\end{center}

We explicitly construct the extended Hermitian Hamiltonian $\hat{\sf H}_{\rm tot}$ in the main text. We begin with an ansatz
\begin{equation}
\hat{\sf H}_{\rm tot} = \hat{I}_{2} \otimes \hat{H}_{S} + \hat{\sigma}_{y} \otimes \hat{V},
\end{equation}
where both $\hat{H}_{S}$ and $\hat{V}$ are required to be Hermitian due to Hermiticity of $\hat{\sf H}_{\rm tot}$. Then, the action of $\hat{\sf H}_{\rm tot}$ on the extended state $\ket{\sf \Psi_{\rm tot}} = \ket{\uparrow} \otimes \ket{\psi_{\rm PT}} + \ket{\downarrow} \otimes \left( \hat{\zeta}^{1/2} \ket{\psi_{\rm PT}} \right)$ is 
\begin{eqnarray}
\hat{\sf H}_{\rm tot} \ket{\sf \Psi_{\rm tot}}
&=& \left[ \hat{I}_{2} \otimes \hat{H}_{S} + \hat{\sigma}_{y} \otimes \hat{V} \right] \left[ \ket{\uparrow} \otimes \ket{\psi_{\rm PT}} + \ket{\downarrow} \otimes \left( \hat{\zeta}^{1/2} \ket{\psi_{\rm PT}} \right) \right] \nonumber \\
&=& \ket{\uparrow} \otimes \left[ \left( \hat{H}_{S} - \ii \hat{V} \hat{\zeta}^{1/2} \right) \ket{\psi_{\rm PT}} \right]
+ \ket{\downarrow} \otimes \left[ \left( \hat{H}_{S} + \ii \hat{V} \hat{\zeta}^{-1/2} \right) \left( \hat{\zeta}^{1/2} \ket{\psi_{\rm PT}} \right) \right].
\end{eqnarray}
The action of $\hat{\sf H}_{\rm tot}$ should be the same as that of the original PT-symmetric Hamiltonian $\hat{H}_{\rm PT}$ when the entire Hilbert space $\mathcal{H}_{2} \otimes \mathcal{H}_{\rm PT}$ is projected by $\hat{\sf P}_{\uparrow} := \hat{P}_{\uparrow} \otimes \hat{I}_{N}$, while the action of $\hat{\sf H}_{\rm tot}$ should be the same as that of the Hamiltonian $\hat{\zeta}^{1/2} \hat{H}_{\rm PT} \hat{\zeta}^{-1/2}$ for the PT-symmetric counterpart when $\mathcal{H}_{2} \otimes \mathcal{H}_{\rm PT}$ is projected by $\hat{\sf P}_{\downarrow} := \hat{P}_{\downarrow} \otimes \hat{I}_{N}$. Therefore, $\hat{H}_{S}$ and $\hat{V}$ satisfy
\begin{equation}
\hat{H}_{S} - \ii \hat{V} \hat{\zeta}^{1/2} = \hat{H}_{\rm PT},~
\hat{H}_{S} + \ii \hat{V} \hat{\zeta}^{-1/2} = \hat{\zeta}^{1/2} \hat{H}_{\rm PT} \hat{\zeta}^{-1/2}.
\end{equation}
Solving the above equations, $\hat{H}_{S}$ and $\hat{V}$ are obtained as
\begin{eqnarray}
\hat{H}_{S} &=& \left( \hat{H}_{\rm PT} \hat{\zeta}^{-1/2} + \hat{\zeta}^{1/2} \hat{H}_{\rm PT} \right) \left( \hat{\zeta}^{1/2} + \hat{\zeta}^{-1/2} \right)^{-1},\\
\hat{V} &=& \ii \left( \hat{H}_{\rm PT} - \hat{\zeta}^{1/2}\,\hat{H}_{\rm PT}\,\hat{\zeta}^{-1/2} \right) \left( \hat{\zeta}^{1/2} + \hat{\zeta}^{-1/2} \right)^{-1},
\end{eqnarray}
both of which are indeed Hermitian. In fact, recalling that $\hat{h} := \hat{\eta}^{1/2} \hat{H}_{\rm PT} \hat{\eta}^{-1/2}$ is Hermitian \cite{Most-02-a} and $\left( \hat{\zeta}^{1/2} + \hat{\zeta}^{-1/2} \right)^{-1} = c^{-1} \hat{\eta}^{-1} \hat{\zeta}^{1/2}$, the above formulas are expressed as
\begin{eqnarray}
\hat{H}_{S} &=& c^{-1} \hat{\eta}^{-1/2} \left( \hat{h} + \hat{\zeta}^{1/2}\,\hat{h}\,\hat{\zeta}^{1/2} \right) \hat{\eta}^{-1/2},\\
\hat{V} &=& \ii\,c^{-1} \left[ \left( \hat{\eta}^{-1/2}\,\hat{h}\,\hat{\eta}^{-1/2} \right) \hat{\zeta}^{1/2} - \hat{\zeta}^{1/2} \left( \hat{\eta}^{-1/2}\,\hat{h}\,\hat{\eta}^{-1/2} \right) \right].
\end{eqnarray}
It is straightforward to confirm the Hermiticity of these operators since $\hat{\eta},\,\hat{\zeta}$ and $\hat{h}$ are all Hermitian operators and $c$ is a real number. Note that the ``interaction" $\hat{H}_{I} := \hat{\sigma}_{y} \otimes \hat{V}$ vanishes if $\hat{H}_{\rm PT}$ is Hermitian. In particular, when the PT-symmetric system consists of two levels ($N=2$), $\hat{\zeta}^{1/2}$ reduces to $\hat{\eta}$ and
\begin{equation}
\hat{H}_{S}^{\rm \left( two-level \right)} = c^{-1} \left( \hat{H}_{\rm PT} \hat{\eta}^{-1} + \hat{\eta} \hat{H}_{\rm PT} \right),~
\hat{V}^{\rm \left( two-level \right)} = \ii\,c^{-1} \left( \hat{H}_{\rm PT} - \hat{H}_{\rm PT}^{\dag} \right).
\end{equation}
We thus reproduce the results of Ref. \cite{GS-08}.

\bigskip
\begin{center}
\textbf{Derivation of Eq.~(11)}
\end{center}

We derive Eq.~(11) in the main text which shows the time evolution of the distinguishability for the two-level system $\hat{H}_{\rm PT}
= s \left( \hat{\sigma}_{x} + \ii a\,\hat{\sigma}_{z} \right)$. The time evolution operator of this system is 
\begin{equation}
\hat{U}_{\rm PT}
= e^{-\ii \hat{H}_{\rm PT} t}
= \frac{1}{\sqrt{1-a^{2}}} \left( \begin{array}{@{\,}cc@{\,}} 
	\sqrt{1-a^{2}} \cos \theta + a \sin \theta 
	& -\ii \sin \theta \\ 
	- \ii \sin \theta 
	& \sqrt{1-a^{2}} \cos \theta - a \sin \theta \\
	\end{array} \right),
\end{equation}
where $\theta := \sqrt{1-a^{2}} st$. We thus have the time-evolved state $\ket{\uparrow \left( t \right)}$ ($\ket{\downarrow \left( t \right)}$) starting from the initial state $\ket{\uparrow}$ ($\ket{\downarrow}$):
\begin{eqnarray}
\ket{\uparrow \left( t \right)}
&=& \frac{\hat{U}_{\rm PT} \ket{\uparrow}}{\| \hat{U}_{\rm PT} \ket{\uparrow} \|}
= \frac{1}{\sqrt{N_{\uparrow}}} \left( \begin{array}{@{\,}c@{\,}} 
	\sqrt{1-a^{2}} \cos \theta + a \sin \theta \\ 
	- \ii \sin \theta \\
	\end{array} \right),~~
	N_{\uparrow} = 1-a^{2}\cos 2 \theta + a \sqrt{1-a^{2}} \sin 2 \theta, \\
\ket{\downarrow \left( t \right)}
&=& \frac{\hat{U}_{\rm PT} \ket{\downarrow}}{\| \hat{U}_{\rm PT} \ket{\downarrow} \|}
= \frac{1}{\sqrt{N_{\downarrow}}} \left( \begin{array}{@{\,}c@{\,}} 
	- \ii \sin \theta \\ 
	\sqrt{1-a^{2}} \cos \theta - a \sin \theta \\
	\end{array} \right),~~
	N_{\downarrow} = 1-a^{2}\cos 2 \theta - a \sqrt{1-a^{2}} \sin 2 \theta.
\end{eqnarray}
Hence the distinguishability between them is calculated to be
\begin{eqnarray}
D \left( \ket{\uparrow \left( t \right)},\,\ket{\downarrow \left( t \right)} \right)
&=& \frac{1}{2}\,\mathrm{tr} \left|\,\ket{\uparrow \left( t \right)} \bra{\uparrow \left( t \right)} - \ket{\downarrow \left( t \right)} \bra{\downarrow \left( t \right)}\,\right| \nonumber  \\
&=& \sqrt{1- \left| \braket{ \uparrow \left( t \right) | \downarrow \left( t \right) } \right|^{2}}
= \left[ 1 + \left( \frac{2a \sin^{2} \theta}{1-a^{2}} \right)^{2} \right]^{-1/2}.
\end{eqnarray}

\bigskip
\begin{center}
\textbf{Criticality at the exceptional point in the PT-symmetric optics}
\end{center}

We present a detailed discussion on the criticality of the information flow under the PT-symmetric periodic potential at the exceptional point
\begin{equation}
V_{\rm PT}^{\left( {\rm EP} \right)} \left( x \right) 
= V_{0} \left[\,\cos \left(\frac{2\pi x}{a}\right) + \ii \sin \left(\frac{2\pi x}{a}\right) \right]
= V_{0}\,\exp \left(\frac{2\pi \ii x}{a}\right).
\end{equation}
Here, Ref.~\cite{Graefe-11} provides the exact solution of the Schr\"odinger equation for this potential. In particular, we take as our input a Gaussian profile of the form 
\begin{equation}
\psi \left( x, 0 \right)
= \exp \left[ - \left( \frac{x}{w} \right)^{2} + \ii k_{0} x \right],
\end{equation}
where $k_{0}$ and $w$ represent the offset and width of the initial beam. In this setup, linear growth of the norm $\left| \psi \left( x, z \right) \right|$ with $z$ emerges in the case of $k_{0} = -1$ \cite{Graefe-11}.

\begin{figure}[H]
\centering
\includegraphics[width=96mm]{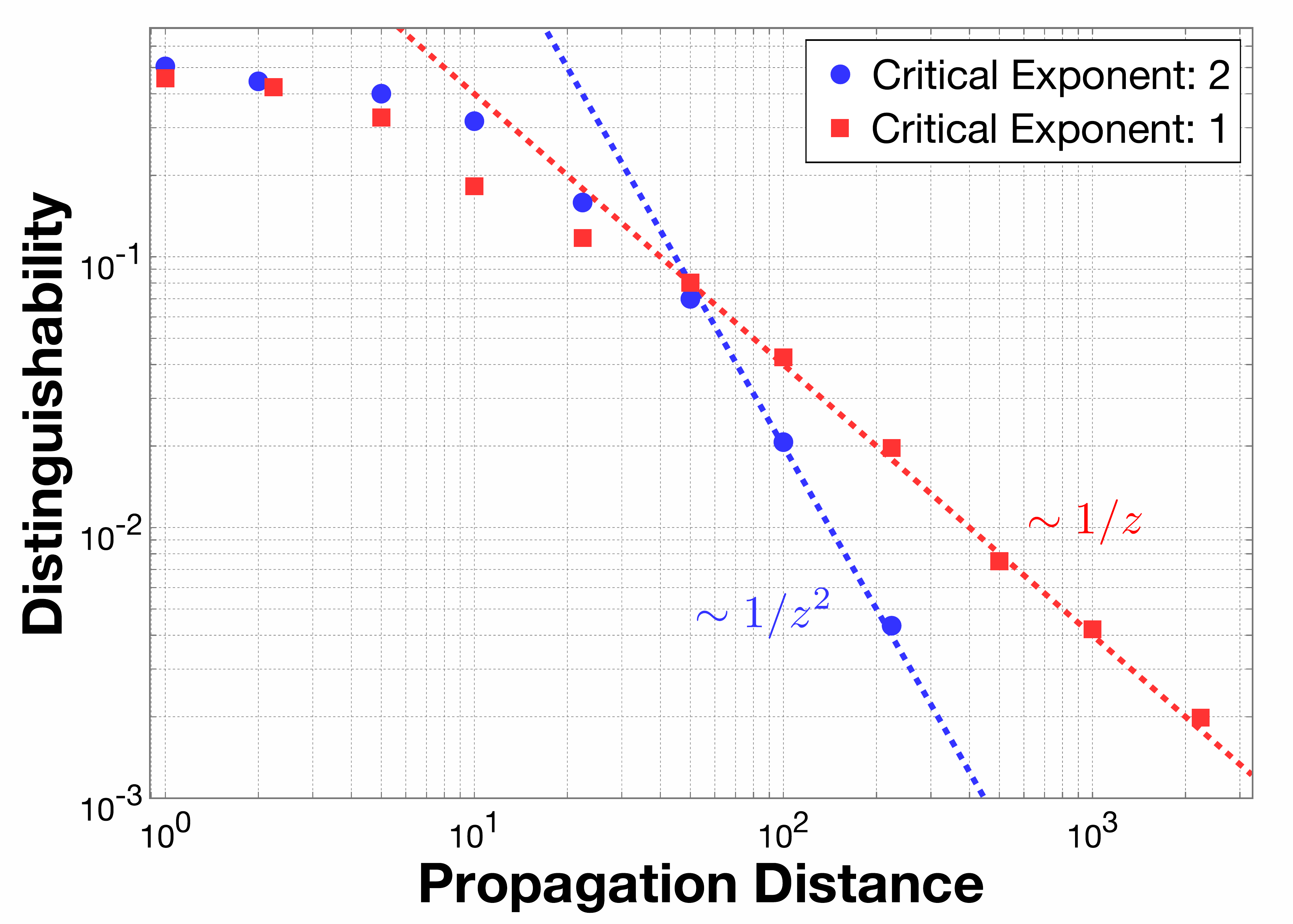} 
\caption{Critical behavior of the distinguishability at the exceptional point. If we choose two initial wavefunctions with different centers, the distinguishability asymptotically behaves as $\left| D - D_{\infty} \right| \sim 1/z^{2}$ for a sufficiently long propagation distance (blue circle points). On the other hand, if we choose two initial wavefunctions with different widths, the distinguishability asymptotically behaves as $D \sim 1/z$ for a sufficiently long propagation distance (red square points).}
	\label{fig: optics}
\end{figure}

We can have two critical exponents of the distinguishability depending on the initial states. On the one hand, if the initial states are chosen to be those with different centers
\begin{equation}
\psi_{1} \left( x, 0 \right) = \exp \left[ - \left( \frac{x - 10}{6 \pi} \right)^{2} + \ii \left( {-1} \right) x \right],~~
\psi_{2} \left( x, 0 \right) = \exp \left[ - \left( \frac{x + {10}}{6 \pi} \right)^{2} + \ii \left( {-1} \right) x \right],
\end{equation}
we have the power-law behavior $\left| D \left( z \right) - D \left( z =\infty \right) \right| \sim 1/z^{2}$ \,(Fig.~\ref{fig: optics}). In this case, the two states do not decay into the same state ($D \left( z =\infty \right) = {0.430531} \neq 0$). On the other hand, if the initial states are chosen to be those with different widths
\begin{equation}
\psi_{1} \left( x, 0 \right) = \exp \left[ - \left( \frac{x}{6 \pi} \right)^{2} + \ii \left( {-1} \right) x \right],~~
\psi_{2} \left( x, 0 \right) = \exp \left[ - \left( \frac{x}{3 \pi} \right)^{2} + \ii \left( {-1} \right) x \right],
\end{equation}
we have the power-law behavior $D \left( z \right) \sim 1/z$ \,(Fig.~\ref{fig: optics}).

\begin{thebibliography}{}
\bibitem{Remark-InnerProduct}
We do not employ a preferentially selected inner product defined in the PT-unbroken phase, with which the dynamics described by $\hat{H}_{\rm PT}$ is equivalent to (or indistinguishable from) the unitary dynamics of a closed system \cite{Brody-16}.
\bibitem{Remark-RelativeEntropy}
If the trace distance is replaced with the quantum relative entropy \cite{BLP-10}, the obtained measure is also contractive under CPTP maps and can be interpreted as information flow. However, it cannot distinguish between pure states since the quantum relative entropy between pure states is either zero or infinite. The classical relative entropy between expectation values of an observable remains finite unlike its quantum counterpart. Nevertheless, this measure explicitly depends on a specific choice of an observable and there exists arbitrariness in its definition.
\bibitem{Remark-Broken-ShortTime}
The distinguishability may increase just after the outset of time evolution for appropriately chosen initial states.
\bibitem{Remark-Weisskopf-Wigner}
It is worth noting that in the Weisskopf-Wigner theory of spontaneous emission irreversibility emerges in infinite and continuum ancillary electromagnetic modes, while atoms undergo Rabi oscillations for a single electromagnetic mode.
\end{thebibliography}
\end{document}